\documentclass[12pt,preprint]{aastex}

\shorttitle{Massive Close Binary HD~115071}
\shortauthors{Penny et al.}

\begin{document}

\received{2002 January 29}
\accepted{}

\title{Tomographic Separation of Composite Spectra. IX. \\ 
The Massive Close Binary HD~115071} 

\author{Laura R. Penny\altaffilmark{1}}
\affil{Department of Physics and Astronomy \\
 College of Charleston \\
 Charleston, SC 29424; \\
 pennyl@cofc.edu}

\author{Douglas R. Gies\altaffilmark{2}}
\affil{Center for High Angular Resolution Astronomy and \\
 Department of Physics and Astronomy\\
 Georgia State University, Atlanta, GA 30303; \\
 gies@chara.gsu.edu}

\author{John H. Wise\altaffilmark{3}}
\affil{School of Physics \\
 Georgia Institute of Technology \\
 Atlanta, GA 30332; \\
 jwise@astro.psu.edu}

\author{D. J. Stickland, C. Lloyd}
\affil{Rutherford Appleton Laboratory \\
Chilton, Didcot, Oxon, OX11 0QX, United Kingdom; \\
ds@astro1.bnsc.rl.ac.uk, cxl@ast.star.rl.ac.uk} 

\altaffiltext{1}{Guest Observer,
Complejo Astronomico El Leoncito (CASLEO), 
San Juan, Argentina} 

\altaffiltext{2}{Guest Observer,
Mount Stromlo and Siding Springs Observatories, Australia}

\altaffiltext{3}{Current address: 
Department of Astronomy and Astrophysics,
Pennsylvania State University, 
532 Davey Laboratory, 
University Park, PA 16802} 

\slugcomment{Submitted to ApJ.}
\paperid{55468}


\begin{abstract}
We present the first orbital elements for the massive close binary, 
HD~115071, a double-lined spectroscopic 
binary in a circular orbit with a period of $2.73135 \pm 0.00003$ 
days.  The orbital semiamplitudes indicate a mass ratio of 
$M_2/M_1 = 0.58 \pm 0.02$ and yet the stars have similar 
luminosities.   We used a Doppler tomography algorithm to 
reconstruct the individual component optical spectra, 
and we applied well known criteria to 
arrive at classifications of O9.5~V and B0.2~III for the 
primary and secondary, respectively.   We present models 
of the {\it Hipparcos} light curve of the ellipsoidal 
variations caused by the tidal distortion of the secondary, 
and the best fit model for a Roche-filling secondary 
occurs for an inclination of $i=48\fdg7 \pm 2\fdg1$.   
The resulting masses are  $11.6\pm1.1 M_\odot$ and $6.7\pm 0.7 M_\odot$
for the primary and secondary, respectively, so that both 
stars are very overluminous for their mass.   
The system is one of only a few known semi-detached, 
Algol-type binaries that contain O-stars.  We suggest 
that the binary has recently emerged from extensive  
mass transfer (possibly through a delayed contact and 
common envelope process). 
\end{abstract}



\section{Introduction}                              

The hot, massive star, HD~115071 (V961~Cen, LS~2998, HIP~64737), 
is found in the sky close to the open cluster, Stock~16 \citep{tur85}, 
and is classified as O9.5~V by \citet{hou75} and B0.5~Vn 
by \citet{gar77}.  The star is not a known visual binary 
\citep{mas98} but early measurements by spectroscopists 
indicated it is radial velocity variable and a
probable spectroscopic binary \citep{cru74,con77}.  
The proof of its binary 
nature came relatively recently in studies by \citet{pen96} and 
\citet{how97}.  Both papers presented a cross-correlation 
analysis of a single, high dispersion, UV spectrum made 
with the {\it International Ultraviolet Explorer Satellite}
({\it IUE}) that demonstrated that the system is in fact 
a double-lined binary.   \citet{sti01} measured the 
radial velocities of the components in this spectrum and
proposed an orbital period of 2.73126~d based upon a 
light curve constructed from {\it Hipparcos} photometry.  
\citet{llo01} present a model of the light curve, and 
they argue that the system has evolved through Case~A 
mass transfer (commencing during core H burning of the donor star). 

The details and outcomes of Roche lobe overflow (RLOF) in massive 
binaries are still subjects of considerable debate \citep{wel01}, 
and thus, the orbital and physical parameters of a system like 
HD~115071 are of great interest.  Here we present the first
double-lined orbital solution for the binary (\S3) based upon 
new high quality optical spectra.  We apply a version of 
the Doppler tomography algorithm (which we have used to good 
effect with UV spectra in prior papers in this series) 
to reconstruct the individual spectra of both components, 
from which we determine the spectral classifications, 
projected rotational velocities, and flux ratio (\S4). 
We also present a light curve analysis constrained by 
the spectroscopic results that allows us to estimate  
the stellar masses (\S5).  These masses are much lower 
than expected, and we discuss the evolutionary implications 
in \S6.  


\section{Observations and Reductions}               
  
Our spectra were obtained in two observing runs at different sites. 
The first set was obtained with the 2.15-m telescope of the 
Complejo Astronomico El Leoncito (CASLEO) and REOSC echelle spectrograph
(on loan from the Institut d'Astrophysique, Universite de Liege, Belgium) 
during the period 1997 March 19 -- 28.  The REOSC spectrograph uses 
an echelle grating with 70 grooves~mm$^{-1}$ and blazed at 226434 \AA 
~together with a cross disperser grating of 400 grooves~mm$^{-1}$ 
blazed at 4000 \AA .  The detector was a TEK $1024\times 1024$ 
CCD with 24$\mu$m square pixels used with a gain of 1.98 e$^-$/ADU 
(read noise of 7.4 e$^-$).  We used a 200$\mu$m slit that 
corresponds to $2\arcsec$ on the sky.  This arrangement produced 
an echellogram from which we extracted 23 orders, spanning the range
from 3575 to 5700 \AA ~with a resolving power of 
$\lambda / \Delta \lambda = 13000$.   We usually obtained 3 exposures 
of 660~s duration that were later co-added in software to 
improve the S/N ($\approx 150$ per pixel in the better exposed 
portions of the spectrum).   Numerous bias, flat field, dark, and Th-Ar
comparison images were obtained each night.    

Our second observing run took place at the 74-inch telescope 
at Mount Stromlo Observatory over the period 1998 April 6 -- 14. 
These spectra were made with the coude spectrograph using grating C
(600 grooves per mm, blazed at 12500\AA ~in first order)
in third order with a BG12 order sorting filter.  The detector was a
SITe CCD (D14) with 15 $\mu$m square pixels in a $4096\times 2048$
format.   This arrangement produced single order spectra 
that covered the range 3804 -- 4220 \AA ~with a
reciprocal dispersion of 0.10 \AA ~per pixel and a resolution
element of 0.30 \AA ~FWHM ($\lambda / \Delta \lambda = 13400$).   
Exposure times were usually 45 minutes, and the final spectra 
have a typical S/N = 160 per pixel in the continuum.
 
The spectra were reduced
using standard routines in IRAF\footnote{IRAF is distributed by the
National Optical Astronomy Observatories, which is operated by
the Association of Universities for Research in Astronomy, Inc.,
under cooperative agreement with the National Science Foundation.}.
The MSO single-order spectra were extracted, calibrated, and 
flux rectified with the task {\it doslit}.
The CASLEO echelle spectra were traced, extracted, and wavelength 
calibrated using the task {\it doecslit}, and the extracted orders 
were rectified to a unit continuum by fitting a high order spline 
function to line-free regions (using the task {\it continuum}). 
Finally the individual orders were linked together with the 
task {\it scombine}.   Small amplitude irregularities related to the
fitting of the echelle blaze function were evident in the 
continuum, and the same residual pattern was seen in all spectra made on a
given night.  We were able to remove most of the pattern 
by dividing the target spectrum by a correction spectrum formed from 
spectra of B-star, $\tau$~Sco, which was also observed each night.  
The correction spectrum was a smoothed version of the particular
night's $\tau$~Sco spectrum divided by a global average 
representation of this star's stellar spectrum.  
The spectra from each run were then collected and  
transformed onto their respective heliocentric wavelength grids.


\section{Radial Velocities and Orbital Elements}    

Our procedure for measuring radial velocities in {\it IUE} spectra 
\citep{pen97} involves fitting Gaussians to the cross-correlation 
functions of the target spectrum with a narrow-lined reference spectrum. 
The optical spectra we consider here have many fewer stellar lines
and much better S/N than the {\it IUE} spectra, so we revised 
our techniques accordingly.  First, we fit each absorption feature 
separately rather than fitting the entire spectrum through 
one cross-correlation measurement.  Secondly, we made the fit of
the composite profiles using spectral templates rather than  
Gaussian functions (since the lines have shapes dominated by 
linear Stark broadening or rotational broadening and since some
lines may contain weak blends).   The templates
were formed from spectra we obtained during each run of the 
star, HD~57682 (O9~IV; \citet{wal72}).  This star is a reasonable  
match in classification to both components in HD~115071 (\S4), 
but has narrower lines  ($V \sin i = 33$ km~s$^{-1}$; \citet{pen96}). 
The radial velocity of this star was measured by parabolic 
fitting of the line cores for lines in the list of \citet{bol78}, 
and we found an average radial velocity of $25.0 \pm 0.5$ and 
$26.0 \pm 1.1$ km~s$^{-1}$ from the CASLEO and MSO spectra, 
respectively.  The averaged template spectra from each run 
were shifted by these values to place them in the rest frame.  
Next, we artificially broadened each template spectrum 
by convolution with a rotational broadening function 
to produce profiles that matched the spectral components
of HD~115071 in the best separated quadrature spectra.  
We also used these resolved profiles to estimate the 
line depth ratio between the components.   Once these 
fitting parameters were set, we determined the radial velocities
of each component for a given line by a least-squares fit of
the observed profile with the coaddition of the two template profiles 
shifted in wavelength to obtain the best match.  This approach 
provided good fits of the observed profiles for all but 
two cases (HJD 2,450,529.792 and 2,450,531.755) 
where the line depth ratio appeared to be reversed. 

We used this technique to measure radial velocities
for the strongest lines in the spectrum, specifically
\ion{H}{1} $\lambda\lambda  3835$, 3889, 3970, 4101, 4340, 4861, 
\ion{He}{1} $\lambda\lambda 3819$, 4009, 4026, 4121, 4143, 4387, 4471, 4921, 5015, 
\ion{He}{2} $\lambda 4686$, and \ion{Si}{4} $\lambda 4089$.
There was no evidence of systematic line-to-line 
differences in the radial velocity measurements, and so 
no line specific corrections were applied. 
The radial velocities from all the available lines were averaged 
together after deletion of any very discrepant measurements. 
Finally, we made small adjustments to these averages based 
on measurements of the strong interstellar \ion{Ca}{2} 
$\lambda\lambda 3933, 3968$ lines.   
An interstellar spectrum was formed by extracting the
mean spectrum in the immediate vicinity of each interstellar
absorption line.  
(We made Gaussian fits of the interstellar \ion{Ca}{2} profiles 
in the extracted spectra, and we found the radial velocity 
was $-17.0\pm 0.2$ and $-16.0 \pm 0.2$ km~s$^{-1}$ for the
mean CASLEO and MSO spectra, respectively.)  
We then cross correlated this spectrum with
each individual spectrum to measure any small deviations in
our wavelength calibration (generally $<3$~km~s$^{-1}$),
and these small corrections were applied to the mean velocities. 
Table~1 lists the 
heliocentric dates of mid-observation, orbital phase, 
and for each component, the mean radial velocity, the standard 
deviation of the mean, the observed minus calculated residual 
from the orbital fit, and the number of lines used in the mean. 
Table~1 also gives the radial velocities from the single {\it IUE} 
spectrum measured by \citet{sti01} (adjusted for the ISM velocity 
on the MSO system).  

\placetable{tab1}      

\citet{sti01} and \citet{llo01} found that the {\it Hipparcos} 
light curve was best fit with a double sine, ellipsoidal 
variation for an orbital period $P=2.73126 \pm 0.00009$~d.  
We found that this period also agreed reasonably well 
with our radial velocity data.  We used the non-linear, least-squares 
fitting program of \citet{mor74} to solve for the period and 
other orbital elements for the primary (the more luminous and 
massive star) and secondary components separately, and this 
yielded period estimates of $2.73149 \pm 0.00007$ and 
$2.73138 \pm 0.00015$~d, respectively.  We made one additional 
calculation of the period by dividing the difference between 
the best fit time of the {\it Hipparcos} photometry maximum and
our spectroscopically determined time of quadrature by the closest integral 
number of cycles, and this led to a period of $2.73130 \pm 0.00004$~d. 
We adopted the error weighted mean of these three estimates for 
our working value of the period, $P=2.73135 \pm 0.00003$~d. 

We fixed this period and then fit for the remaining orbital 
elements independently for both components.  The fitted 
epoch of primary maximum velocity, $T_0$, was the same within 
errors for both solutions, and so we applied the mean value 
to fits of both components.  Eccentric solutions produced 
estimates of eccentricity consistent with a value of zero, 
and our final solutions in Table~2 assume circular motion. 
The observed and calculated radial velocity curves appear 
in Figure~1.  The only major discrepancies occur in the 
{\it IUE} measurements (not used in the solution), 
both of which are $\approx 38$ km~s$^{-1}$
above the predicted curve.  Note that in the case of the primary, 
the {\it IUE} velocity falls well above the maximum for the 
entire curve, so the mismatch cannot be due to an incorrect 
orbital phase for example.   The systematic difference may be related 
to line formation at different heights in an expanding atmosphere 
or orbital motion about a distant, unseen, tertiary star.

\placetable{tab2}      

\placefigure{fig1}     


\section{Tomographic Reconstruction}                

We used the Doppler tomography algorithm described by 
\citet{bag94} to reconstruct the individual primary and 
secondary spectra independently from the CASLEO and MSO 
spectra.  We took the radial velocity shifts for each component 
from the orbital solutions in Table~2, then the reconstruction 
was run for 50 iterations with a gain of 0.8 (the results 
are insensitive to both parameters).  The reconstructed 
spectra are plotted in in Figure 2 in a format similar 
to that used in the spectral atlas of \citet{wal90}.  The 
reconstructions from the MSO spectra are shown just above those from 
the CASLEO spectra (in the short wavelength portion of Fig.~2), 
and there is good agreement between these two sets of spectra.

\placefigure{fig2}     

We compared the reconstructed spectra with the spectrum 
standards in the atlas of \citet{wal90} to determine the 
spectral classifications of the components.   
The strengths of the \ion{He}{1} $\lambda\lambda 4026,4143,4387$ 
lines relative to those of \ion{He}{2} $\lambda\lambda 4200, 4541$
are all consistent with a spectral type of O9.5 for the primary. 
The ratio of the \ion{Si}{4} $\lambda\lambda 4088,4116$ lines
to the nearby \ion{He}{1} $\lambda\lambda 4121,4143$ features 
indicate a main sequence class, as does the relatively strong 
\ion{He}{2} $\lambda 4686$ to \ion{He}{1} $\lambda 4713$ ratio. 
Thus, we classify the primary as type 
O9.5~V, and we compare its spectrum in Figure~2 to that of 
HD~93027, which is given as the standard of this class  
in \citet{wal90}. 

The secondary, on the other hand, has features indicating a 
cooler temperature and later type.  The ratio of  
\ion{Si}{3} $\lambda 4552$ to \ion{Si}{4} $\lambda 4088$
has a good match in the interpolated type B0.2 introduced 
by \citet{wal90}.   The relative strength of the  
\ion{Si}{4} $\lambda\lambda 4088,4116$ lines compared to the 
neighboring \ion{He}{1} $\lambda\lambda 4121,4143$ features
clearly leads to a luminosity class III.  Figure~2 
illustrates the good agreement between the spectrum of the 
secondary and that of HD~108639 that \citet{wal90} use as 
a standard for type B0.2~III.  The \ion{C}{3} $\lambda\lambda 
4070, 4650$ blends appear to be somewhat weaker in the 
secondary's spectrum than in the standard spectrum (evidence, 
perhaps, of CNO-processed gas in the secondary's photosphere). 

The two spectral standards, HD~93027 and HD~108639, provided 
us with the means to estimate the visual flux ratio, 
$r=F_2/F_1$, by matching the line depths in the reconstructed
spectra with those in the standards.   This was done by 
aligning the reconstructed and standard spectra, 
adjusting for differences in the placement of the continuum,
Gaussian smoothing of the spectra to eliminate 
differences in projected rotational velocity and instrumental 
broadening, and then finding a best fit line ratio that 
allocates a proportion of flux to each component to 
best match the line depths.   We found $r=1.04 \pm 0.06$ and 
$1.08\pm 0.08$ for the MSO and CASLEO reconstructions, respectively. 

Finally, we used the profiles in the reconstructed spectra 
to estimate the projected rotational velocities of the 
components.  We focused on the \ion{Si}{4} $\lambda 4088$
profile for this purpose since it represents the strongest 
metallic line (intrinsically narrow) in the range covered 
by the MSO spectra.   Our procedure involved calculating a 
grid of rotational broadening functions for a linear 
limb darkening law \citep{wad85,gra92} and then  
convolving an observed narrow-lined spectrum with 
these broadening functions.   We compared the spectral reconstructions 
from the MSO spectra with broadened versions of MSO spectra of
the narrow-lined stars HD~53682 (O9~IV) and $\tau$~Sco (B0.2~V).   
The best fitting profile matches were made with 
$V\sin i = 101\pm 10$ and $132\pm 15$ km~s$^{-1}$ for the 
primary and secondary, respectively.   These agree 
within errors with estimates from the {\it IUE}
observation \citep{pen96,how97,llo01}. 


\section{Light Curve Analysis and Masses}           

\citet{llo01} presented an analysis of the {\it Hipparcos}
light curve \citep{per97}, and here we update their work 
by restricting a number of the fitting parameters based upon 
the new spectroscopic results.  
We used the light curve synthesis code GENSYN \citep{moc72} 
to produce model $V$-band differential light curves (almost identical to
differential {\it Hipparcos} $Hp$ magnitudes for hot stars). 
The orbital parameters were taken from the spectroscopic solution,
and the physical parameters were estimated from the spectral
classifications of the stars.   
We first estimated the stellar temperature 
and gravity according to the spectral classification
calibration of \citet{how89} for the primary 
($T_{{\rm eff}~1} = 32$~kK, $\log g_1 = 3.9$), 
and for the secondary, we used data for comparable stars 
in the compilation of \citet{und82} 
($T_{{\rm eff}~2} = 29$~kK, $\log g_2 = 3.6$).
We then determined the physical fluxes and limb darkening coefficients 
from tables in \citet{kur94} and \citet{wad85}, respectively.
We also used the Kurucz flux models to transform our observed 
flux ratio based upon the relative line depths into a 
$V$-band flux ratio \citep{pen97}.
The MSO spectra are centered at 4009~\AA, and the transformation 
yields a $V$-band flux ratio, $F_2/F_1 = 1.05 \pm 0.06$. 
The comparison of line depths in the CASLEO spectra was made over
the available range in the standard spectrum from \citet{wal90} 
(centered at 4350~\AA ), and the resulting $V$-band flux ratio
is $F_2/F_1 = 1.09 \pm 0.06$.  We used the average value, 
$F_2/F_1 = 1.07 \pm 0.06$, in the light curve analysis.  
The theoretical and observed flux ratios together 
yield an approximate estimate of
the ratio of stellar radii, $R_2/R_1 = 1.12 \pm 0.03$. 
Each trial run of GENSYN was set by two independent
parameters, the system inclination $i$ and secondary's 
radius relative to the critical Roche-filling case 
(with the primary radius set so that the orbital average 
flux ratio matched the observed flux ratio).

The observed light curve (Fig.~3) is a double-sine wave 
caused by tidal distortion in the stars.  Since the 
stars have similar radii but the secondary has a much lower mass (\S3),
the secondary must be much closer to filling its critical 
Roche radius, so that the tidal generation of the light curve 
is due mainly to the distortion of the secondary. 
The amplitude of the photometric variation is proportional 
to the degree of tidal distortion (how close the secondary 
comes to filling its Roche volume) and to the sine of the 
inclination (maximal effect for $i=90^\circ$). 
Our first fit of the light curve assumed that the secondary 
completely fills its Roche volume, so this solution corresponds 
to the case of minimum inclination (and maximum masses).   
The best fit for this semi-detached configuration is made with 
an inclination, $i=48\fdg7 \pm 2\fdg1$, 
and this fit is shown as the solid line in Figure~3.   
The error in the inclination results from two sources, 
the variation in the $\chi^2$ residuals of the fit with parameter $i$ 
and the change in the solution introduced by the uncertainty
in the flux ratio.   The root mean square of the residuals from 
the best fit is 0.019~mag, which is approximately $1.7\times$
larger than the errors quoted in the {\it Hipparcos} catalog, 
and so some other kind of photometric variation may exist that  
is unrelated to orbital phase. 

\placefigure{fig3}     

Note that it is possible to obtain fits 
with a lower inclination if the flux ratio constraint is 
abandoned.  For example, we found that if we assumed a 
contact configuration in which both stars fill their Roche volumes, 
then we could make a satisfactory fit of the light curve with 
$i=38^\circ$.  However, we rule out this model because it predicts a
flux ratio, $F_2/F_1 = 0.52$, that is far below the limits 
established from the spectra of the components. 

Models with a smaller secondary and less tidal distortion require 
a higher inclination to match the observations (yielding lower masses), 
but these solutions are less satisfactory for two reasons.  
First, higher inclination solutions generally yield light curves with 
less ellipsoidal variation but some evidence of eclipses. 
We show one example in Figure~3 for an inclination  
$i=60^\circ$ and a secondary volume radius of     
$R_2/R_\odot = 5.6$ ($\approx 90\%$ of the critical Roche radius).
Eclipses as subtle as those shown in Figure~3 are probably 
not ruled out by the {\it Hipparcos} photometry, but 
models with $i > 62^\circ$ show eclipses that are clearly 
inconsistent with the {\it Hipparcos} light curve.  
Secondly, the projected rotational velocities predicted 
by underfilling models with synchronous rotation are much smaller 
than the observed values.  All the known binaries containing 
O-stars with periods this short have circular orbits 
\citep{mas98}, and we expect that such close systems 
have attained synchronous rotation as well \citep{cla97}.
The predicted projected rotational velocities are 
$V\sin i = 92$ and 109 km~s$^{-1}$ for the 
primary and secondary, respectively, in the Roche-filling model, 
in agreement within errors with the observed values (\S4). 
However, the match is worse in higher inclination models 
($V\sin i = 81$ and 94 km~s$^{-1}$, respectively, for the 
$i=60^\circ$ model illustrated in Fig.~3).   Thus, we prefer 
the secondary Roche-filling model, and we list in Table~3 the 
corresponding stellar parameters.  
The system absolute magnitude in this model 
is $M_V = -4.57$, and, for $V=7.94$ and $E(B-V)=0.50$ 
\citep{tur85}, we estimate a distance of $1.5\pm0.2$~kpc
(smaller than but comparable to the distance of 1.9~kpc for 
the cluster Stock~16; \citet{tur85}). 

\placetable{tab3}      


\section{Discussion}                                

The first striking result from our analysis is the very low 
mass we find for both components.  The stars have temperatures 
and luminosities that are associated with masses of 18 and 
$15 M_\odot$ for the primary and secondary, respectively, in the 
single star evolutionary tracks calculated by \citet{sch92}.
(These estimates would be slightly reduced using evolutionary
models that include rotation; \citet{heg00}, \citet{mey00}.) 
The secondary, in particular, has a luminosity characteristic 
of a star more than twice as massive than we find (Table 3).    

The second remarkable fact is that the secondary star 
has a spectral classification indicating it has 
evolved away from the main sequence.   Thus, HD~115071 presents 
the classical ``Algol paradox'' that the lower mass 
component is the more evolved one, and we suggest the same 
solution of the paradox holds here as well, i.e., that the 
evolved component was originally the more massive object, 
but suffered significant mass transfer to its neighbor. 

There are only a small number of O-stars that are known 
to be members of interacting binaries, and we compare 
in Table~4 the properties of the components in HD~115071 
with those of the four other known semi-detached binaries 
that contain O-type stars \citep{hil87,hh98}.  We excluded from 
this list contact or over-contact systems and those binaries 
in which both components are evolved \citep{van98}.  All the 
systems in Table~4 share a number of common properties: the mass 
donor appears as an evolved star, the donor star is overluminous 
for its mass, the donor fills its Roche volume, and the mass
gainer is a late O-type, main sequence star.
It is remarkable that all the donor stars have comparable 
luminosity, $\log L/L_\odot \approx 4.5$, despite their wide 
range in mass and radius.   Evolutionary models generally 
predict that the post-RLOF luminosity of the donor is 
comparable to its zero age main sequence (ZAMS) 
luminosity \citep{van98,wel01}, 
and so these donors probably began life as B0~V stars 
with masses in the range 14 -- $20 M_\odot$.  Since the 
donors were originally the more massive component, the 
gainers were probably also B-type stars that were 
promoted to their current O-type status through mass transfer. 
It is also curious that no semi-detached systems are known with 
primaries earlier than type O8~V.  Either this stage is extremely 
rapid in more massive systems or the donor stars take on a 
different appearance than they do in Algol-type systems 
(perhaps as a O-star plus Wolf-Rayet star binary; \citet{van98}).

\placetable{tab4}      

Evolutionary models give us some guidance about the initial 
masses in HD~115071.  \citet{del94} give a relationship between 
the final, post-RLOF mass and the initial ZAMS mass, and this 
yields an estimate of $14.8 M_\odot$ for the initial mass 
of the donor star.  If we further assume that 50\% of the 
donor's mass loss was accreted by gainer and the rest lost 
from the system \citep{meu89,del94}, then the original total 
mass was $22.4 M_\odot$ and the original gainer mass was $7.6 M_\odot$. 
Thus, the system probably began with a relatively low mass ratio, 
$M_g/M_d \approx 0.5$. 

The theoretical models of binary evolution by \citet{wel01} offer 
some guidance in the interpretation of our results.   
\citet{wel01} describe the evolution of several very close systems 
that begin RLOF during core H-burning (Case A).   Their 
models suggest that a mass reversal similar to what we find 
in HD~115071 can occur in Case A, but the resulting systems 
generally have a much wider orbit and more extreme mass ratio. 
Another possibility is that the system began RLOF after 
completion of core H burning (Case B).  The initial period 
would have been much larger, but the system then shrunk to 
its current dimensions during a common envelope phase 
in which the donor's envelope would have been ejected from 
the system.  This would explain the current mass and 
luminosity of the donor star, but it does not account for 
the huge overluminosity of the contemporary primary star, 
which is the most overluminous star of any of the gainers 
in Table 4.   \citet{wel01} point out one other hybrid 
scheme they call ``delayed contact'' in which mass transfer 
begins conservatively until the donor develops a convective 
envelope and the binary enters the common envelope stage.  
This scenario would explain the observed low mass of the 
donor and the short orbital period, and the overluminosity 
of the gainer would result from compression and/or mixing 
related to mass accretion.    

The best fit of the light curve suggests that the secondary 
donor star is Roche-filling, and so the system may 
still be experiencing active mass transfer.  Observations 
of any H$\alpha$ emission \citep{tha97} or IR excess \citep{geh95} 
would provide valuable clues about the mass loss and/or 
mass transfer processes that might be occurring presently 
in this exceptional binary system. 


\acknowledgments

We thank the staffs of CASLEO and MSO for their assistance 
in making these observations.   We are grateful to Norbert 
Langer for comments on the evolutionary state of the binary.  
Institutional support for L.R.P.
has been provided from the College of Charleston School
of Sciences and Mathematics.  Additional support for L.R.P.\ was
provided by the South Carolina NASA Space Grant Program and
NSF grant AST-9528506.
Institutional support for D.R.G. has been provided from the GSU College
of Arts and Sciences and from the Research Program Enhancement
fund of the Board of Regents of the University System of Georgia,
administered through the GSU Office of the Vice President
for Research.  We gratefully acknowledge all this support.





\clearpage

\begin{figure}
\caption{
The radial velocity measurements 
({\it primary -- filled circles; secondary -- open circles}) 
and orbital solution ({\it solid lines}) plotted against
orbital phase.  Phase zero corresponds to the time of primary 
maximum radial velocity. The two plus marks show the {\it IUE}
measurements that were not used in the solution.}
\label{fig1}
\end{figure}

\begin{figure}
\caption{
A comparison of the reconstructed 
MSO spectra ({\it above}) and CASLEO spectra ({\it below}) 
of the primary and secondary with spectra of the same classifications 
from \citet{wal90}.  All the spectra were Gaussian smoothed 
to a nominal resolution of 1.2 \AA ~FWHM for consistent line broadening.}
\label{fig2}
\end{figure}

\begin{figure}
\caption{
The {\it Hipparcos} light curve plotted against spectroscopic 
orbital phase. The solid line shows the predicted curve for a 
secondary Roche-filling model with $i=48\fdg7$, while the dashed line 
represents the prediction for an under-filling model with $i=60^\circ$.}
\label{fig3}
\end{figure}


\clearpage

\begin{deluxetable}{cccccccccc}
\tabletypesize{\scriptsize}
\tablewidth{0pt}
\tablenum{1}
\tablecaption{Radial Velocity Measurements \label{tab1}}
\tablehead{
\colhead{HJD}             &
\colhead{Orbital}         &
\colhead{$V_1$}           &
\colhead{$\sigma_1$}      &
\colhead{$(O-C)_1$}       &
\colhead{}                &
\colhead{$V_2$}           &
\colhead{$\sigma_2$}      &
\colhead{$(O-C)_2$}       &
\colhead{}                \\
\colhead{(-2,400,000)} &
\colhead{Phase}           &
\colhead{(km s$^{-1}$)}   &
\colhead{(km s$^{-1}$)}   &
\colhead{(km s$^{-1}$)}   &
\colhead{$n_1$}           &
\colhead{(km s$^{-1}$)}   &
\colhead{(km s$^{-1}$)}   &
\colhead{(km s$^{-1}$)}   &
\colhead{$n_2$}           }
\scriptsize
\startdata
44487.472 & 0.921 &       \phs 104.4 &\nodata&  \phs 34.0 &\nodata &      $-151.6$ &\nodata&  \phs 41.2 &\nodata \\
50526.858 & 0.057 &    \phn\phs 80.5 & 6.2 & \phn\phs 3.6 &     17 &      $-189.6$ & 2.2 &    \phs 14.5 &     12 \\
50527.780 & 0.394 &         $-117.3$ & 3.5 &   \phn$-4.2$ &     17 &    \phs 131.3 & 6.8 & \phs\phn 5.3 &     17 \\
50528.845 & 0.784 &  \phn\phn $-1.8$ & 5.5 & \phn\phs 1.2 &     17 &  \phn $-47.6$ & 5.1 &    \phs 17.7 &     16 \\
50529.792 & 0.131 &    \phn\phs 52.3 & 2.7 & \phn\phs 3.7 &     13 &      $-160.3$ & 2.1 &  \phn $-5.4$ &     14 \\
50530.777 & 0.491 &         $-122.8$ & 7.1 &    \phs 13.5 &     17 &    \phs 182.1 & 7.8 &    \phs 15.6 &     16 \\
50531.755 & 0.849 &    \phn\phs 32.6 & 6.3 &   \phn$-5.4$ &     17 &      $-148.0$ & 4.6 &      $-11.5$ &     17 \\
50532.783 & 0.226 &  \phn\phn $-0.3$ & 6.0 & \phn\phs 9.4 &     16 &  \phn $-28.2$ & 7.7 &    \phs 25.5 &     16 \\
50533.754 & 0.581 &         $-125.5$ & 4.4 &   \phn$-3.0$ &     17 &    \phs 159.1 & 8.4 &    \phs 16.8 &     17 \\
50535.766 & 0.318 &     \phn $-82.8$ & 4.7 &      $-10.8$ &     15 & \phn\phs 43.4 & 3.1 &      $-11.2$ &     12 \\
50910.125 & 0.378 &     \phn $-98.5$ & 1.7 & \phn\phs 7.3 & \phn 9 &    \phs 107.6 & 2.3 &   \phn$-5.7$ &     10 \\
50911.086 & 0.730 &     \phn $-22.1$ & 2.6 &    \phs 18.1 & \phn 9 &  \phn $-31.6$ & 5.6 &      $-31.0$ &     10 \\
50911.171 & 0.761 &     \phn $-29.4$ & 0.4 &      $-10.4$ & \phn 6 &  \phn $-34.3$ & 5.4 & \phs\phn 3.2 &     10 \\
50916.213 & 0.607 &         $-119.0$ & 4.1 &   \phn$-6.4$ &     10 &    \phs 122.8 & 5.6 &   \phn$-2.3$ &     10 \\
50916.990 & 0.891 &    \phs\phn 58.8 & 2.5 &   \phn$-0.2$ & \phn 9 &      $-166.8$ & 2.4 & \phs\phn 6.2 & \phn 9 \\
50917.116 & 0.938 &    \phs\phn 73.9 & 4.6 &   \phn$-1.5$ &     10 &      $-204.6$ & 3.5 &   \phn$-3.1$ & \phn 9 \\
50917.257 & 0.989 &    \phs\phn 79.8 & 5.4 &   \phn$-3.7$ &     10 &      $-220.4$ & 4.4 &   \phn$-4.8$ &     10 \\
50917.971 & 0.250 &      \phn$-26.8$ & 1.8 &   \phn$-0.1$ & \phn 8 &  \phn $-28.0$ & 4.9 &   \phn$-3.8$ & \phn 9 \\
50918.112 & 0.302 &      \phn$-59.8$ & 3.8 & \phn\phs 2.0 & \phn 8 & \phn\phs 23.6 & 2.7 &      $-13.4$ & \phn 9 \\
50918.249 & 0.352 &         $-105.3$ & 3.2 &      $-13.0$ &     10 & \phn\phs 77.5 & 2.4 &      $-12.4$ & \phn 9 \\
\enddata
\end{deluxetable}

\newpage

\begin{deluxetable}{lr}
\tablewidth{0pc}
\tablenum{2}
\tablecaption{Circular Orbital Elements  \label{tab2}}
\tablehead{
\colhead{Element} & 
\colhead{Value} }
\startdata
$P$~(days)           \dotfill        & 2.73135 (3) \\
$T_0$ (HJD-2,400,000) \dotfill       & 50734.286 (11) \\
$K_1$ (km s$^{-1}$)    \dotfill      & 110.1 (28) \\
$K_2$ (km s$^{-1}$)     \dotfill     & 191.4 (48) \\
$V_{0~1}$ (km s$^{-1}$) \dotfill     & $-26.4$ (19) \\
$V_{0~2}$ (km s$^{-1}$) \dotfill     & $-24.6$ (32) \\
$m_1$ sin$^{3}i$ ($M_\odot$)\dotfill & 4.94 (40) \\
$m_2$ sin$^{3}i$ ($M_\odot$)\dotfill & 2.84 (26) \\
$a_1$ sin $i$ ($R_\odot$) \dotfill   &  5.94 (15) \\
$a_2$ sin $i$ ($R_\odot$) \dotfill   & 10.32 (26) \\
r.m.s.$_1$ (km s$^{-1}$)  \dotfill   &  8.4 \\
r.m.s.$_2$ (km s$^{-1}$)  \dotfill   & 14.2 \\
\enddata
\tablecomments{Numbers in parentheses give the error in the last digit quoted.}
\end{deluxetable}

\newpage

\begin{deluxetable}{lcc}
\tablewidth{0pc}
\tablenum{3}
\tablecaption{Stellar Properties \label{tab3}}
\tablehead{
\colhead{Property} & 
\colhead{Primary} &
\colhead{Secondary} }
\startdata
Spectral Classification  \dotfill          &  O9.5~V       & B0.2~III \\
Relative flux $F/F_1$(5470\AA )\dotfill    &  1.0          & $1.07\pm0.06$ \\
$V\sin i$ (km s$^{-1}$)        \dotfill    &  $101\pm 10$  & $132\pm 15$ \\
$T_{\rm eff}$ (kK)             \dotfill    &  $32 \pm 2$   & $29 \pm 1.5$  \\
$M/M_\odot$                    \dotfill    &  $11.6\pm1.1$ & $6.7\pm 0.7$ \\
$R/R_\odot$                    \dotfill    &  $6.5\pm 0.2$ & $7.2\pm 0.2$ \\
$\log g$                       \dotfill    &$3.88\pm 0.01$ & $3.55\pm 0.01$ \\ 
$\log L/L_\odot$               \dotfill    &$4.60\pm 0.14$ & $4.52\pm 0.12$   \\
\enddata
\end{deluxetable}

\newpage

\begin{deluxetable}{lcllcrccc}
\tablewidth{0pc}
\tabletypesize{\footnotesize}
\tablenum{4}
\tablecaption{Semi-Detached OB-Star Binaries \label{tab4}}
\tablehead{
\colhead{} & 
\colhead{$P$} &
\colhead{Pri.} &
\colhead{Sec.} &
\colhead{$M_P$} &
\colhead{$M_S$} &
\colhead{} &
\colhead{} &
\colhead{} \\
\colhead{Name} & 
\colhead{(d)} &
\colhead{Type} &
\colhead{Type} &
\colhead{($M_\odot$)} &
\colhead{($M_\odot$)} &
\colhead{$\log L_P/L_\odot$} &
\colhead{$\log L_S/L_\odot$} &
\colhead{Ref.} 
}
\startdata
HD 115071 = V961 Cen       \dotfill & 2.73 & O9.5 V & B0.2 III &  $11.6\pm1.1$ &  $6.7\pm0.7$ &$4.60\pm 0.14$ &$4.52\pm 0.12$& 1 \\
HD 209481 = LZ Cep         \dotfill & 3.07 & O8.5   & O9.5     &  $15.1\pm0.4$ &  $6.3\pm0.2$ &$4.90\pm 0.03$ &$4.65\pm 0.03$& 2 \\
BD$+66^\circ~1521$ = XZ Cep\dotfill & 5.10 & O9.5 V & B1 III   &  $15.8\pm0.4$ &  $6.4\pm0.3$ &$4.58\pm 0.04$ &$4.48\pm 0.03$& 3 \\
HD 106871 = AB Cru         \dotfill & 3.41 & O8 V   & B0.5     &  $19.8\pm1.0$ &  $7.0\pm0.7$ &$5.21\pm 0.03$ &$4.58\pm 0.03$& 4 \\
HD 190967 = V448 Cyg       \dotfill & 6.52 & O9.5 V & B1 II-Ib &  $25.2\pm0.7$ & $14.0\pm0.7$ &$4.54\pm 0.04$ &$4.66\pm 0.06$& 3 \\
\enddata
\tablecomments{References: 
(1) this paper, 
(2) \citet{har98},
(3) \citet{har97},
(4) \citet{lor94}.}
\end{deluxetable}


\clearpage

\setcounter{figure}{0}

\begin{figure}
\plotone{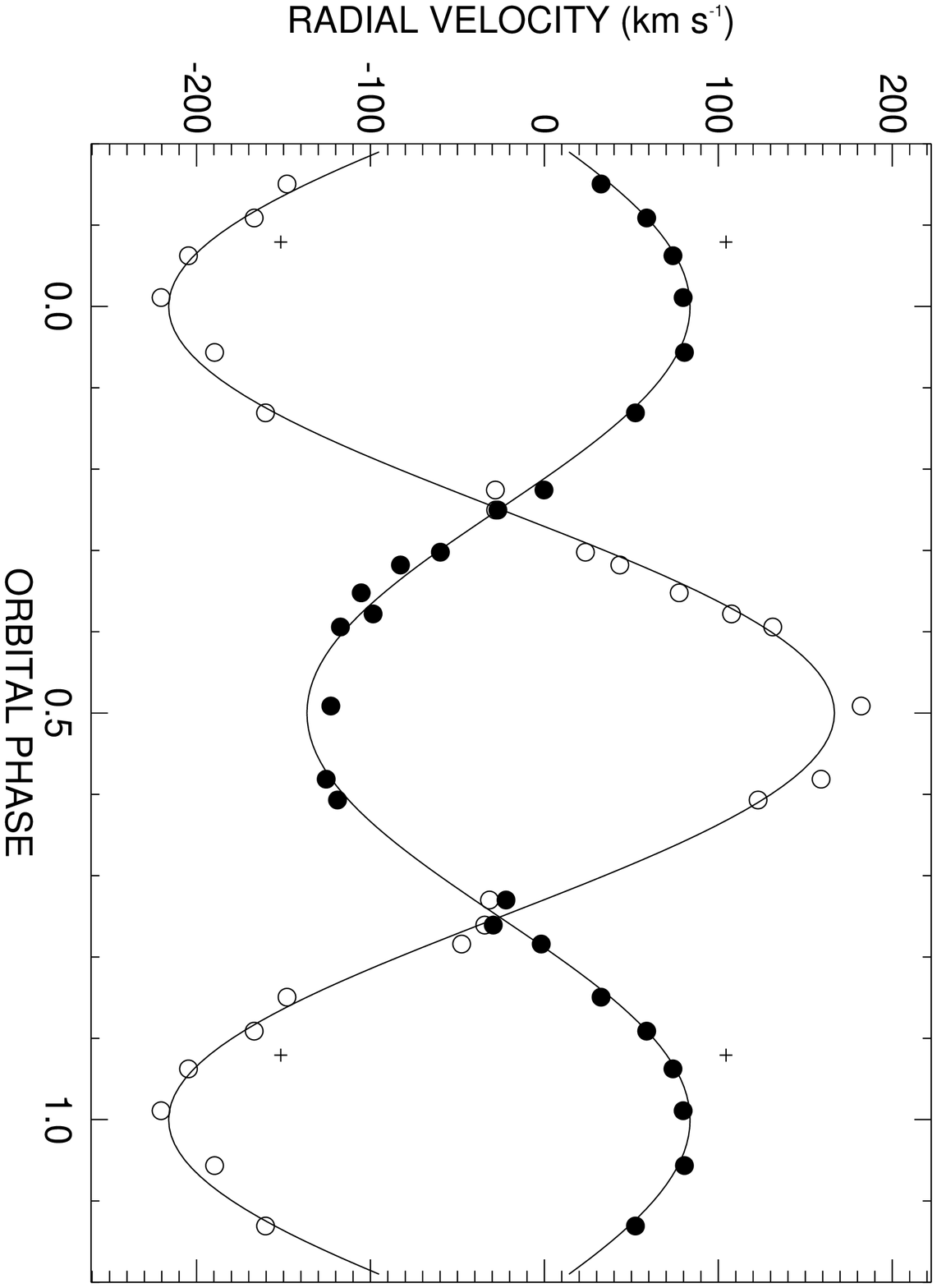}
\caption{}
\end{figure}

\begin{figure}
\plotone{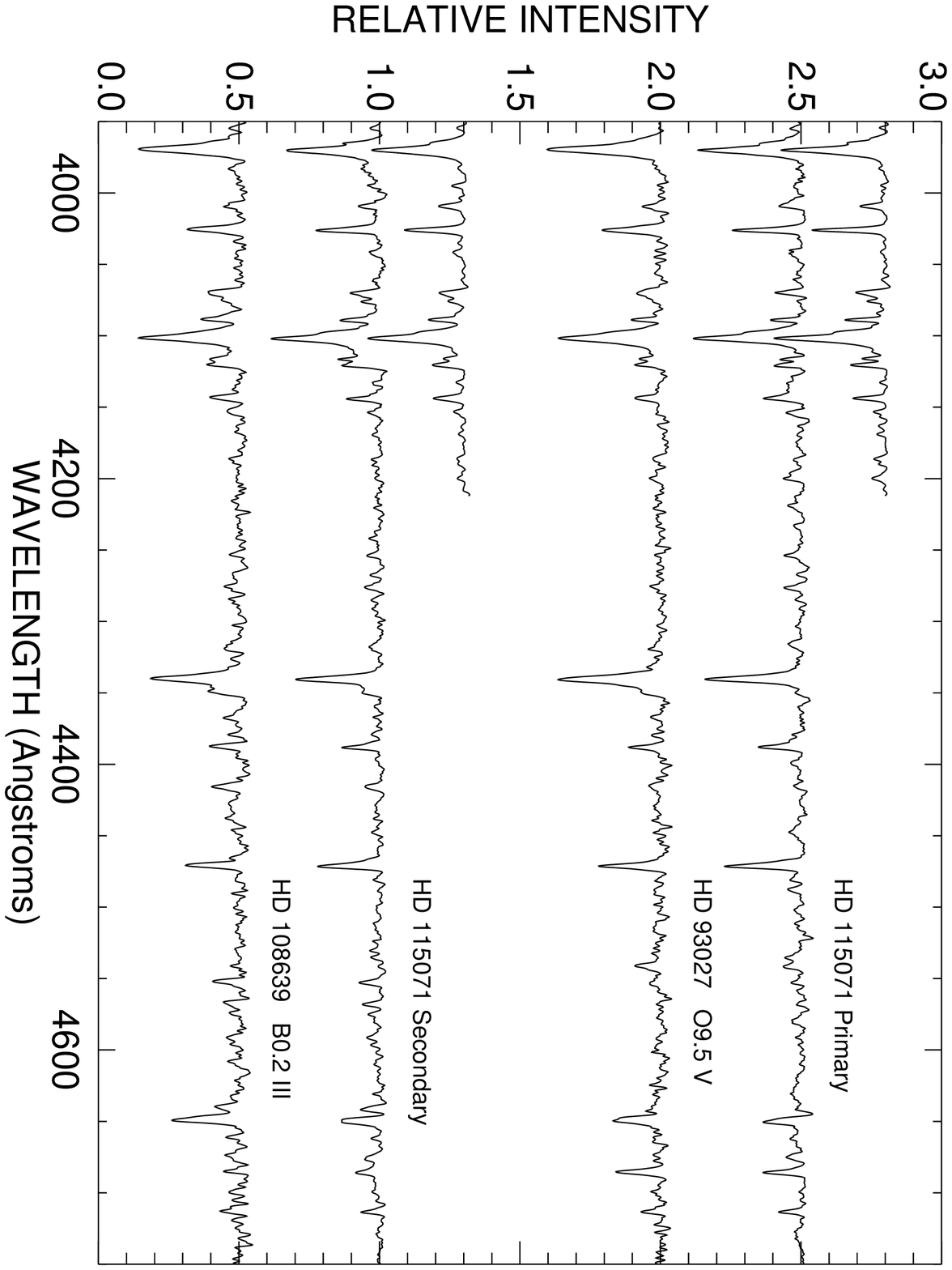}
\caption{}
\end{figure}

\begin{figure}
\plotone{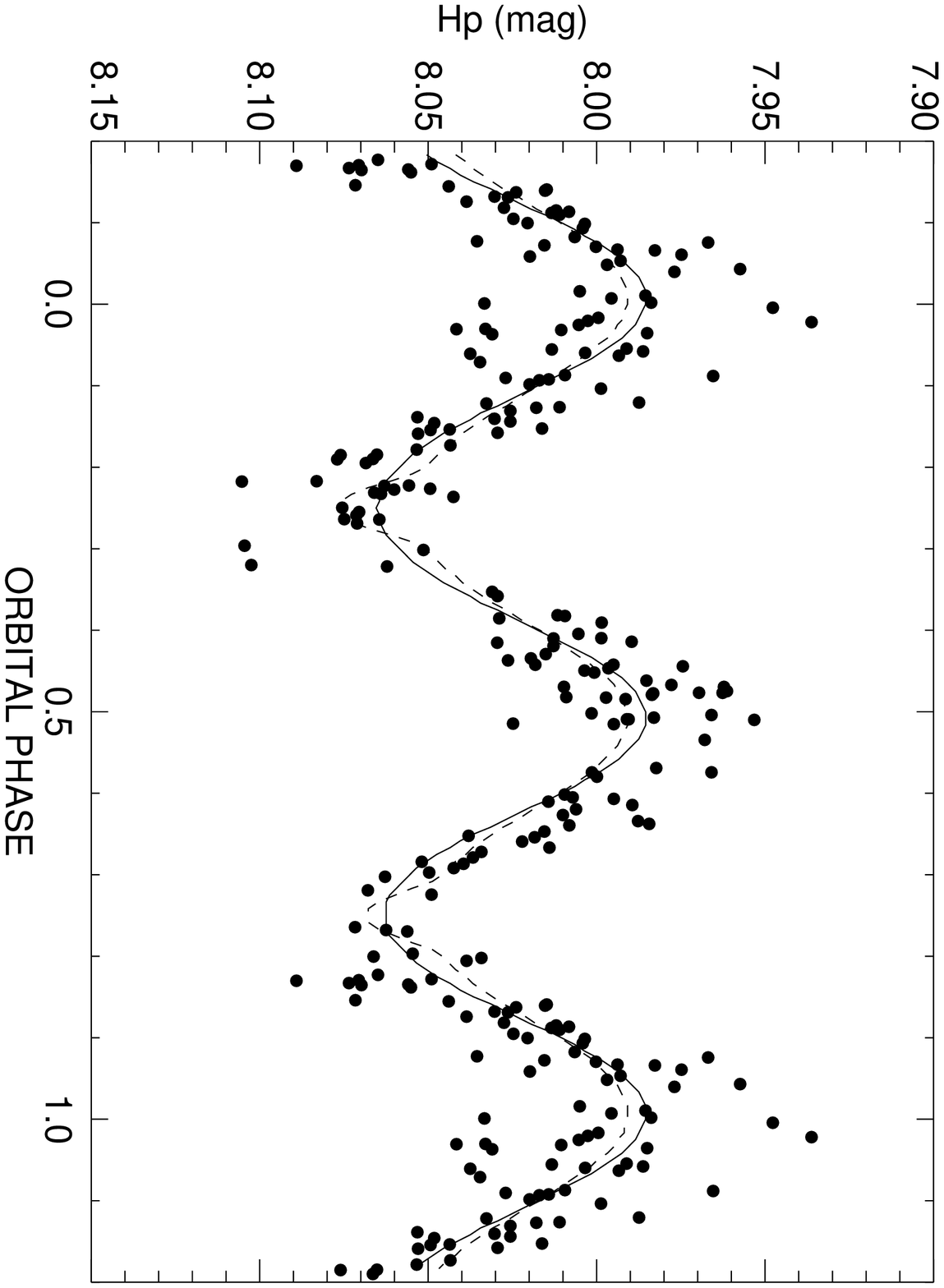}
\caption{}
\end{figure}


\end{document}